\documentclass[%
 reprint,
 amsmath,amssymb,
 aps,
]{revtex4-2}

\usepackage{graphicx}
\usepackage{dcolumn}
\usepackage{bm}

\begin{document}

\title{Spatial resolution, noise and information in the computational-imaging era}

\author{David M. Paganin}
  \email{David.Paganin@monash.edu}
\affiliation{School of Physics and Astronomy, Monash University, Clayton 3800, Australia}

\author{Alexander Kozlov}%
\affiliation{ARC Centre in Advanced Molecular Imaging, School of Physics, The University of Melbourne, Parkville 3010, Australia}

\author{Timur E. Gureyev}
\affiliation{ARC Centre in Advanced Molecular Imaging, School of Physics, The University of Melbourne, Parkville 3010, Australia; School of Physics and Astronomy, Monash University, Clayton 3800, Australia; School of Science and Technology, University of New England, Armidale 2351, Australia; Faculty of Health Science, The University of Sydney, Sydney 2006, Australia}

\date{\today}

\begin{abstract}
Imaging is an important means by which information is gathered regarding the physical world. Spatial resolution and signal-to-noise ratio are underpinning concepts.  There is a paucity of rigorous definitions for these quantities, which are general enough to be useful in a broad range of imaging problems, while being also sufficiently specific to enable precise quantitative evaluation of the relevant properties of imaging systems. This is particularly true for many modern forms of imaging that include digital processing of the acquired imaging data as an integral step leading to final images presented to an end-user. Here, both the well-known historical definitions of spatial resolution and some more recent approaches suitable for many forms of modern computational imaging are discussed. An intrinsic duality of spatial resolution and signal-to-noise exists in almost all types of imaging, with the related uncertainty relationship determining a trade-off between the two quantities. Examples are presented with applications to super-resolution imaging, inline holography and ghost imaging. 
\end{abstract}

\maketitle


\section{Introduction}

Dennis Gabor's landmark paper \cite{1} developed holographic imaging as a two-step process. In the first step an indirect, coded form of the image is recorded, which does not and in general will not bear a direct resemblance to the object that is being imaged.  In the second step, the coded form of the image is decoded, to give the required reconstructed image of the object. This idea of a two-step approach to imaging, namely image recording following by image reconstruction, may be somewhat arbitrarily read as a key moment in the historical development of computational imaging \cite{2}. Hence we speak of Gabor's 1948 paper as dividing the ``before computational imaging'' (BCI) era from the ``after computational imaging'' (ACI) era. 

\subsection{Imaging before the computational-imaging era} 

In the BCI era, optical imaging was typically viewed as equivalent to a one-step process in which images were necessarily {\em directly} recorded, registered or otherwise apprehended. A particularly ancient example is the human visual system. This may be the naked eye, or augmentations thereof using optical telescopes, microscopes, magnifying glasses or corrective lenses. Film-based photography was another milestone of the BCI era. Other advances include the extension of telescopy and microscopy (together with photography) to forms of imaging quanta other than visible light: x-rays, infra-red light, gamma rays, electrons, neutrons, muons etc.  

In all of the above means of imaging, one has a one-step process where optical information is manipulated at the field level by a given optical system, with the resulting image being ``direct'' in the sense that its morphology {\em as recorded} bears a direct spatial resemblance to that of the object being imaged.  The image may be blurred, distorted, scratched, discolored, grainy, blotched or otherwise imperfect, but its morphology still bears a direct relation to the object being imaged. It was in this BCI era, and in this BCI context of what constitutes a necessarily-imperfect image, that elementary concepts of spatial resolution (SR) and signal-to-noise ratio (SNR) were developed.  We briefly consider each concept in turn.   

SNR can be relatively easily defined using elementary statistics, as the ratio of the mean value of the image intensity to its standard deviation within a flat (uniform) patch of the image: 
\begin{equation}
    {\textrm{SNR}}=\frac{\overline{I}(x,y)}{\sqrt{\,\overline{[I(x,y)-\overline{I}(x,y)]^2}}}.
\end{equation}
Here, $I(x,y)$ denotes a two-dimensional measured intensity distribution as a function of Cartesian coordinates $(x,y)$, and an overline denotes ensemble averaging over repeated measurements.  SR is less amenable to elementary definition. A standard example is the Rayleigh criterion \cite{3,4,5,6} for distinguishing the image of two points via a given optical imaging system: see Fig.~1a and the associated caption. Note the strong {\em a priori} knowledge implicit in such a definition, namely the knowledge that the object being imaging consists of two points \cite{4}. 

Toraldo di Francia writes that ``the image of two points, however close to one another, is certainly different from the image of one point'' \cite{4}. In the never-realizable zero-noise limit, one could always determine the separation between two points in an image of these two points, no matter how close together they might be.  From another perspective, if too few photons are collected, the fact that the Rayleigh or related criteria are satisfied will be irrelevant: one will not be able to tell the difference between an image of one point and an image of two points, if only a small number of photons (say, less than five) is recorded. These zero-noise and high-noise limiting cases highlight the connection between resolution and noise.  This link has been remarked upon, both explicitly and implicitly, by many authors in many contexts \cite{8,9,10a,10b,11,12,13,14,15,16,17,18}. These and related studies all conclude the concept of resolution to be meaningless if considered in isolation from noise. As a corollary, which reinforces a point made earlier, we note that noise-independent definitions of resolution, such as the previously-mentioned Rayleigh criterion, implicitly assume zero noise and may therefore be viewed as a limiting case of the statement that noise and resolution are intimately related. We return to this key point in due course.  

\subsection{Rayleigh, Sparrow and Abbe resolution criteria} 

The Rayleigh criterion gives the following well-known estimate for the spatial resolution of an optical system utilizing monochromatic light with wavelength $\lambda$ in vacuum, and angular acceptance $2\theta$ \cite{3,4}: 
\begin{equation}
 {\textrm{SR}}_{\textrm{Rayleigh}}=\frac{1.22\lambda}{\sin\theta}.    
\end{equation} 
Further information on the Rayleigh criterion, together with the related Sparrow and Abbe criteria, is given below. These criteria are formulated differently, and give estimates for SR that depend on important details such as the state of coherence of the ensemble of imaging quanta and the dimensionality of the image, yet all yield an estimate for SR that agrees with ${\textrm{SR}}_{\textrm{Rayleigh}}$ up to multiplicative factors on the order of unity. 

\begin{figure}[h!]
\centering\includegraphics[width=8.5cm]{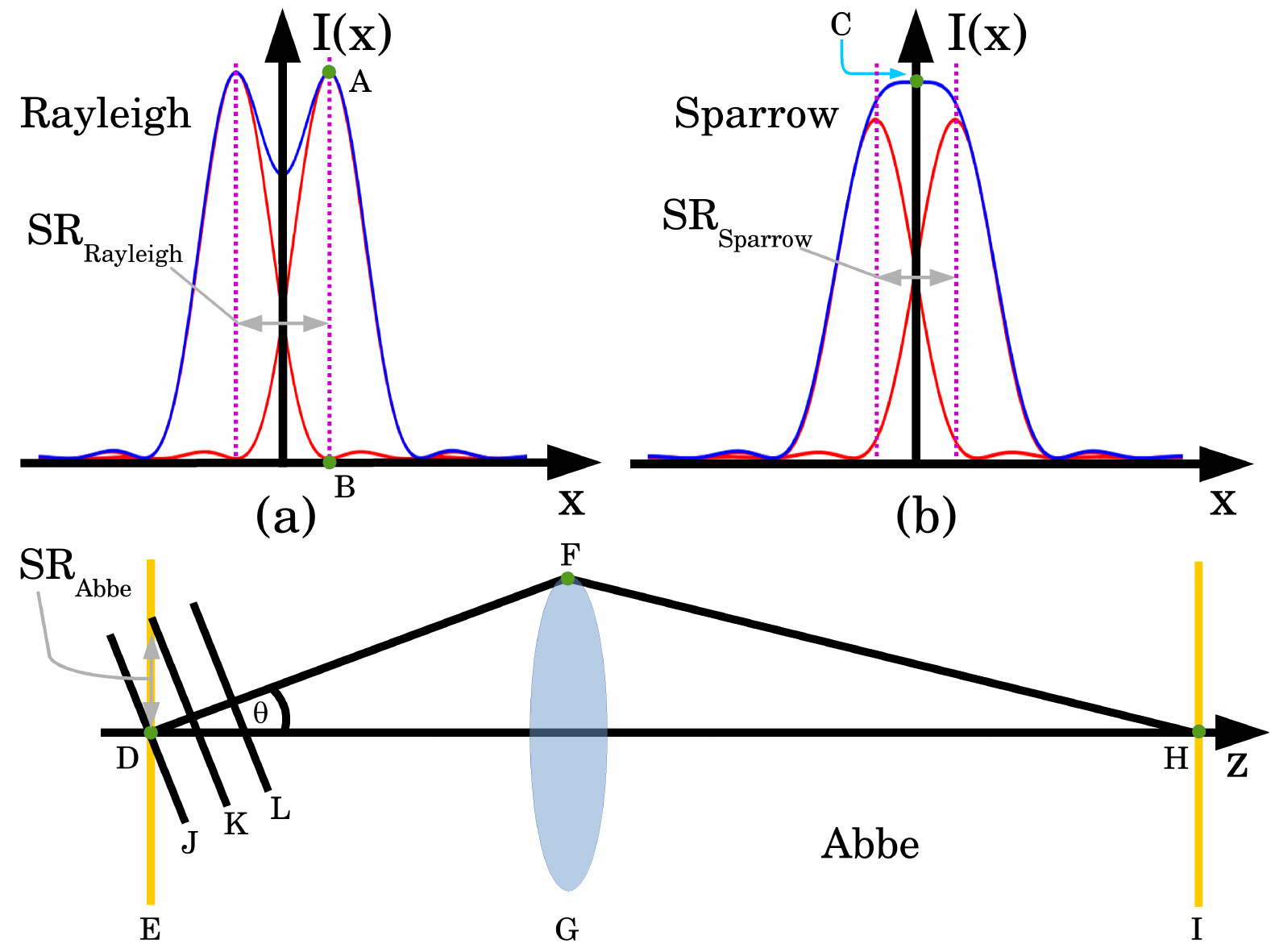}
\caption{RAYLEIGH, SPARROW AND ABBE RESOLUTION CRITERIA. (a) The Rayleigh resolution is based on resolving an image of two objects that are known {\em a priori} to be point-like.  This implies the corresponding intensity image to be two displaced point spread functions (PSFs).  ${\textrm{SR}}_{\textrm{Rayleigh}}$ corresponds to the first zero $B$ of one PSF (left red curve) being co-located with the central maximum $A$ of a second PSF (right red curve), to give the net intensity $I(x)$ shown in blue.  (b) If the two red curves in (a) are brought together until the central dip at $C$ just disappears, according to the criterion that $\partial^2 I(x)/\partial x^2$ is zero at $x=0$, then the individual PSF maxima are considered to be separated by the Sparrow resolution ${\textrm{SR}}_{\textrm{Sparrow}}$. (c) The Abbe resolution ${\textrm{SR}}_{\textrm{Abbe}}$ is the transverse wavelength of the wave-fronts $JKL$, relative to the plane $E$, propagating along the line $DF$ making a maximum angle between the optic axis $Z$ and the lens $G$. }
\end{figure}

The Rayleigh criterion \cite{3,4} states that the SR is given by the width of the first zero of the point spread function (PSF).  Thus, for example, if one has a rotationally-symmetric 2D point spread function $\textrm{PSF}(r)$, where $r$ denotes radius in plane polar coordinates, then SR is the smallest distance that solves $\textrm{PSF}(SR)=0$.  The Rayleigh SR corresponds to the first zero of the PSF associated with a coherently-illuminated circular aperture; in this case SR is equal to the radius of the Airy disc at the center of the PSF.  The Rayleigh criterion, illustrated in Fig.~1a, implies the image of two points to be ``just resolved'' when their separation is such that the central peak of the PSF centered at one point is co-located with the first zero of the PSF centered at the other point.   

The Sparrow criterion \cite{5} gives the point-to-point separation for which the central dip, in the image of the two points, just disappears (see Fig.~1b).  This can be quantified by calculating the smallest non-zero point-separation $B$ such that $\partial^2 I(x,y;B)/\partial x^2 = 0$, where $I(x,y;B)$ is the intensity output by the imaging system when the input consists of two points separated by $B=\textrm{SR}_{\textrm{Sparrow}}$.   

Roughly speaking, the Abbe SR considers the finest resolvable feature, in an imaging system, to be the reciprocal of the highest transverse spatial (Fourier) frequency that is passed by the system \cite{6}.  This may be thought of as the transverse wavelength of the most-highly-tilted plane wave that can be passed through the system, namely a plane wave making an angle of $\theta$ with respect to the optic axis. Thus e.g.~in the optical microscope shown in Fig.~1c, the Abbe SR will be $\lambda/\sin\theta$, again up to factors on the order of unity.  Note that a similar argument appears in a quantum-mechanical context, when considering the ``Heisenberg microscope'' \cite{7}.  

\subsection{Resolution as the width of the PSF}

Both Rayleigh and Sparrow definitions are explicitly based on the PSF, with Abbe SR being implicitly based on the PSF. A closely related, but more general alternative definition of SR is also based on the PSF. In many optical systems the final image can be represented as a convolution of an ``ideal'' image, which can be obtained e.g.~with monochromatic plane-wave illumination and perfect detector (with point-like pixels), with a PSF of the imaging system which accounts for blurring due to polychromaticity, finite source size, detector pixel size, other real-life ``imperfections'', numerical aperture etc. In such cases, and in broad agreement with the SR criteria already mentioned, the SR of the images can usually be equated with the ``width'' of the PSF, where the width can be defined e.g.~as the full width at half-maximum of the PSF, the standard deviation of the PSF: 
\begin{equation}
    \textrm{SR}_{\textrm{PSF}}=\sqrt{\frac{\int |{\bf r}-{\overline{\bf r}}|^2 \textrm{PSF}({\bf r}) \, d{\bf r}}{\int \textrm{PSF}({\bf r}) \, d{\bf r}}}
\end{equation}
or other natural means. As the PSF represents the response (image) produced by the system for point-like input, it is easy to appreciate that the width of the PSF determines the minimal distance between two tightly localized objects at which they can still be resolved in the image. In other words, the width of the PSF is closely related to the Rayleigh and related definitions of the spatial resolution, even though it is possible to devise ``pathological'' cases where the two definitions give very different results. Note also that this approach to SR can be extended to shift-variant cases, in which the images can be obtained by a general linear transformation (not necessarily shift-invariant) of the corresponding ``ideal'' images obtained with monochromatic plane-wave illumination and a perfect detector. In such cases, the role of the point-input response is played by the Green function, $G({\bf r},{\bf r}')$, with the corresponding PSF in the shift-invariant cases being $G({\bf r}-{\bf r}',0)$. In the general shift-variant case, one can define the SR as the spatial average of the width of the Green function, e.g.  
\begin{equation}
    \textrm{SR}_{\textrm{Green}}=\sqrt{\frac{\iint |{\bf r}-{\overline{\bf r}}|^2 G({\bf r},{\bf r}') \, d{\bf r} d{\bf r}'}{\iint G({\bf r},{\bf r}') \, d{\bf r}d{\bf r}'}}.
\end{equation}

\section{Imaging in the computational-imaging era }

In the ACI era, one has a bifurcation, with both direct imaging systems (one-step systems) and indirect imaging systems (two-step systems) enjoying ongoing development \cite{2, 19}. The development of direct imaging systems, again speaking in rather broad terms, was typically seen to strive for progressively higher SR, sensitivity, SNR, magnification etc. Again, all manipulation at the optical field level was considered to be effected by a physical optical system, whose function was to yield a direct image whose SR, SNR etc.~could then be determined. However, something fundamentally new occurred for indirect imaging modalities such as inline holography, off-axis holography, computed tomography, crystallography, ptychography, coherent diffractive imaging, ghost imaging, super-resolution fluorescence microscopy etc.~\cite{2}. In all of these  indirect imaging systems, one has variants of Gabor's idea of imaging as a two-step process, in which one first records data that contain information regarding the object in coded or encrypted form that does not necessarily have a morphology that bears a direct resemblance to the object. This first step may again employ physical-optics elements to manipulate imaging quanta (e.g.~photons, electrons, atoms etc.), to form the encrypted data which may or may not take the form of optical images. One then has a second step, that of reconstruction, in which the previously mentioned data (e.g.~holograms \cite{1}, diffraction patterns \cite{20}, tomographic projections \cite{21}, ghost-imaging bucket measurements \cite{22} etc.) are decoded or decrypted, to compute a reconstruction of the desired object. In such means for indirect imaging, information is manipulated at the field level by the physical optical elements, including the position-sensitive detector, with the subsequent decoding constituting a manipulation of optical information at the digital level via the computer programs which embody the mathematical algorithms performing the image reconstruction. Under this view, one has a hybrid of (a) hardware-optics elements streaming and manipulating imaging quanta at the level of complex field amplitudes, coupled to (b) virtual-optics elements streaming and manipulating image information quanta at the digital-bits level. In such hybrid physical-virtual imaging systems, the computer is an intrinsic part of the imaging system \cite{19,23}.  Typically, the ``old'' criteria for SNR and SR that were formulated in the BCI era, become increasingly unsuitable for such ACI-era imaging systems.  We give some examples below.

\subsection{STORM resolution is not limited by the width of the instrumental PSF} 

Consider a fluorescence-based imaging system in which the instrumental PSF, here termed $\textrm{PSF}_{\textrm{instrument}}$, has a width on the order of the Rayleigh estimate for SR.  Two such spread functions, centered at points $A$ and $B$ on a DNA strand $C$, are shown in Fig.~2a.  If a short-exposure image is taken such that fluorophores at $A$ and $B$ are the only sources of photons during that exposure, the centroids of the well-separated detected instrumental spread functions may each be located with precision significantly better than the widths of these spread functions.  As shown in Fig.~2b, this gives points $A'$ and $B'$.  Repeating the above process, many times, builds up a super-resolution image whose SR is significantly better than that given by the Rayleigh and related criteria.  This idea underpins stochastic optical reconstruction microscopy (STORM) and related techniques \cite{24,25,26}.       

Rather than being dictated by the width of $\textrm{PSF}_{\textrm{instrument}}$, under the Rayleigh and related resolution criteria, this form of super-resolved fluorescence microscopy has a resolution governed by the precision with which the centroid of {\em one} instance of $\textrm{PSF}_{\textrm{instrument}}$ can be located.  The effective computational-imaging spread function is given by the diameter of the red discs in Fig.~2b, rather than the diameter of $\textrm{PSF}_{\textrm{instrument}}$ in Fig.~2a.  
In this and related super-resolution techniques, the Rayleigh criterion is inapplicable. 

\begin{figure}[h!]
\centering\includegraphics[width=8.5cm]{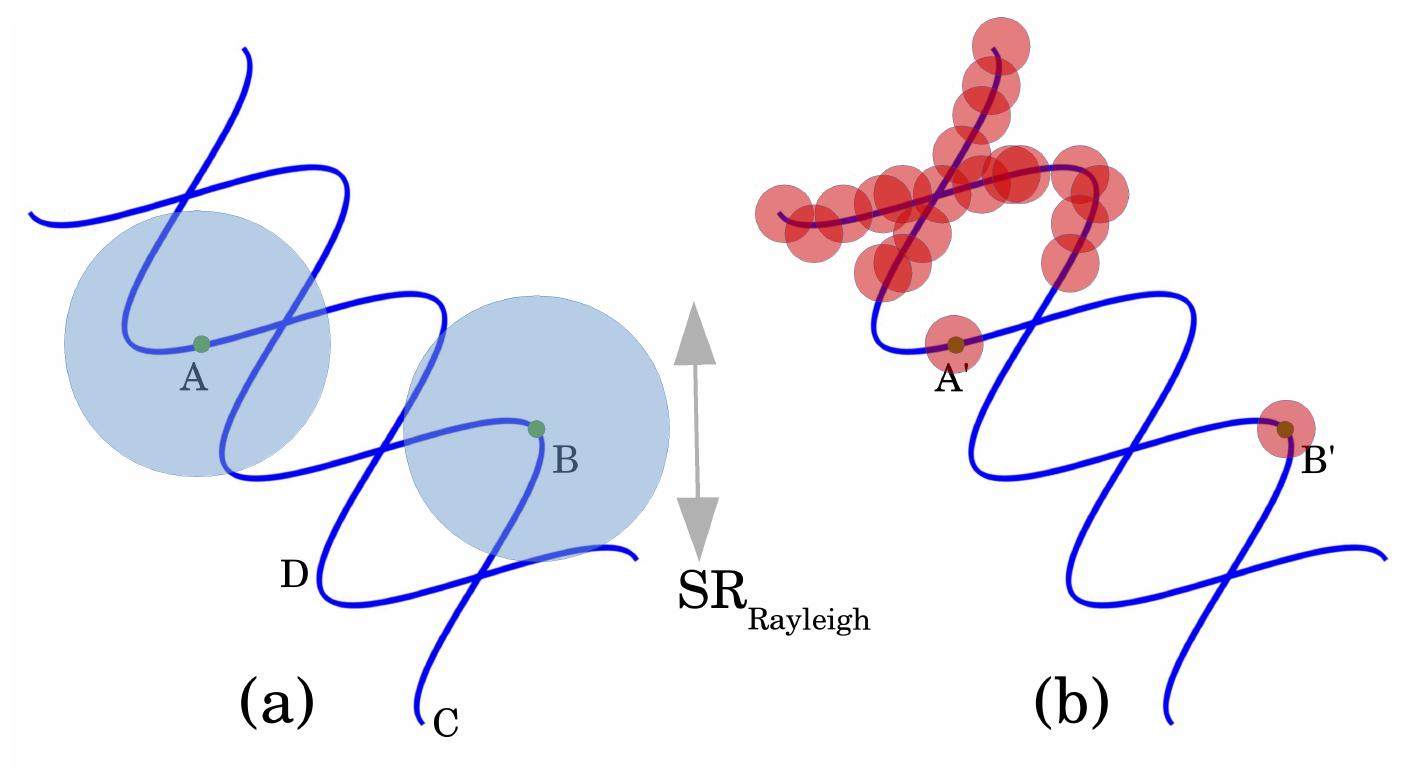}
\caption{SUPER RESOLUTION USING STORM AND RELATED IMAGE TECHNIQUES. (a) In STORM super-resolution fluorescence microscopy \cite{24,25,26}, together with several closely related super-resolution techniques, a short-exposure image of a sample such as a DNA strand receives light from a small number of well-spaced fluorescing points such as $A$ and $B$ on the strands $C$ and $D$ respectively.  The short-exposure image is composed of instrumental point spread functions $\textrm{PSF}_{\textrm{instrument}}$ centered at $A$ and $B$ respectively, as shown by the large blue discs.  (b) The centroids of the discs centered at $A$ and $B$ may be located with a resolution equal to the radius of the red discs shown at $A'$ and $B'$ in the reconstructed STORM image.  The process is repeated many times, as indicated by the series of red discs (only a small number of such discs is shown, for clarity).   }
\end{figure}

\subsection{Ghost-imaging resolution is not limited by the width of the illuminating speckle beam}

Another example, of the inapplicability of the ``old'' criteria for SR and SNR, is given by the case of ghost imaging \cite{22}.  Emerging from the field of quantum optics and initially erroneously believed to be always underpinned by quantum-mechanical ``spooky action at a distance'', the field has rapidly achieved prominence in studies using classical visible light \cite{22}. 

This form of imaging is counter-intuitive.  See Fig.~3, which shows a typical ghost imaging setup, albeit one using x-rays rather than the more-often used visible light, since such a setup recently demonstrated tomographic ghost imaging of optically opaque objects for the first time \cite{27,28}.  In such ghost imaging, photons from an x-ray source pass through a speckle-making mask, leading to a spatially random pattern being measured over the surface of a position-sensitive detector.  A beam-splitter then removes a very small fraction of the photons, which pass through an object and are then recorded by a single-pixel ``bucket'' that merely records the total number of photons falling upon it.  This process is repeated for many different transverse mask positions, and many object orientations.  While no photon that ever passes through the object is ever registered by a position-sensitive detector, and no photons measured by the position sensitive detector ever pass through the object, the {\em correlation} between the two can be used to reconstruct a computational three-dimensional image (tomogram) of the object \cite{22}. 

\begin{figure}[h!]
\centering\includegraphics[width=9cm]{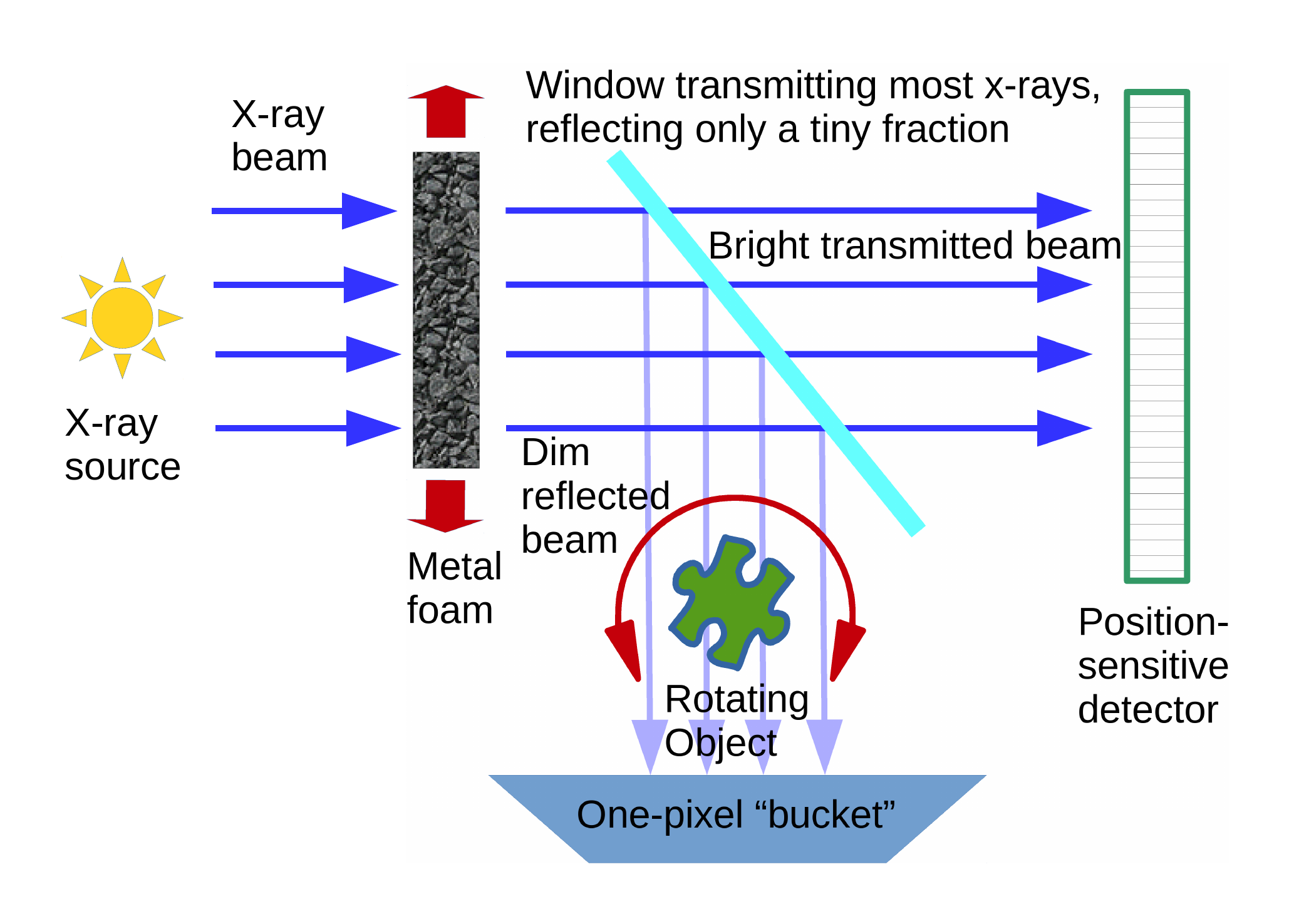}
\caption{TOMOGRAPHIC GHOST IMAGING. A photon source illuminates a spatially random mask, generating a speckle pattern whose intensity distribution is recorded by a position-sensitive detector.  A beam splitter is used to create a weak copy of the speckled beam, which is transmitted through an object, beyond which a so-called bucket detector measures the total number of transmitted photons.  This process is repeated for many mask positions and object orientations.  One then seeks to reconstruct the 3D density of the object via the correlations between the two beams \cite{27,28}.     }
\end{figure}

In ghost imaging, the resolution may be determined via the width of the auto-covariance of an ensemble of spatially random intensity maps of optical speckle \cite{29,30}, and has nothing to do with any lenses that may have been used to manipulate imaging quanta passing through the sample.  Indeed,  sample-scattered quanta are never registered by a position sensitive detector.  Interestingly, resolution is here determined using information derived from photons that never pass through the object, yet which alone contain no information regarding the object.  Again, we see a case from modern computational optics where the classic criteria for SR and SNR break down. 

\subsection{Image noise, partial coherence,  irradiation dose}

Related to the distinction between direct and computational imaging discussed above, is the increased relevance in the ACI era of the effect that the stochastic and, ultimately, quantum nature of imaging, has on the SR, SNR and other imaging characteristics. Just as image processing was already ``covertly'' widespread in the BCI era, for example in interferogram analysis, the everyday significance of the quantum nature of light was also more relevant to everyday visual experience of humans than could be appreciated at the time. As far as image processing is concerned, the human brain intrinsically performs a form of Fourier analysis on every image that is formed on the retina of our eyes. Indeed, specialized areas of the primary visual cortex automatically detect and separate features of the image that extend predominantly in a certain direction, i.e.~it analyses the angular spectrum of the image \cite{31}. On the quantum side of things, human eyes can detect a faint signal containing as little as three photons \cite{31}. Such low-light images are automatically ``denoised'', processed and analyzed by the brain. Evidently, low-SNR image detection and analysis constitutes an inherent part of the human visual system. There is an interesting historical irony here: current exciting trends in artificial neural networks for image processing \cite{32} were pre-dated, by as many hundreds of millennia as {\em homo sapiens} and earlier humans have existed, by the natural-neural-network plus optical-sensor that is the human eye--brain system. Returning to the main argument, we note that, with the advent of the BCI era, the quantum and stochastic nature of light has become a subject of explicit analysis, which was first embedded in statistical optics theories and later in the quantum optics. Most relevant to our present discussion is the effect of these quantum and stochastic behaviors on SR and SNR.  

In fact, the theme of imaging at low-light levels is very much a contemporary phenomenon, which is currently also the subject of intense competition between manufacturers of photographic cameras for smart phones. Consumers have noticed that, as a result of rapid technological development over recent years, even budget-level smart phones are now often capable of producing excellent photographs in dim conditions. The leading manufacturers in this area are now competing for ever-better imaging under low-light conditions, such as at night in the open or in poorly lit indoor environments. These efforts parallel sustained effort to image very faint astronomical objects, together with ongoing work in military optics. However, unlike astronomical and military applications where it may be possible to increase the collected photon flux by increasing the size of optical elements (hence, the development of giant mirrors and antenna arrays), the quest for the perfect image in the smart phone camera arena is increasingly being undertaken in the realm of image processing algorithms. As is typical for the ACI era, increasing sophistication and focus upon the second (reconstruction) stage of imaging is the area of most R\&D effort and swiftest progress. However, the fundamental laws of optical physics can be neither ignored nor bent by means of particularly-efficient image processing algorithms. Especially under low-light conditions, where the stochastic and quantum nature of light becomes more prominent, it is important to keep the corresponding laws and principles of optics in perspective.  

\section{Noise--resolution duality}

We have already emphasized that SR cannot be meaningfully considered without reference to SNR and {\em vice versa}. It is only in the high-SNR limit that SR can be treated independently from SNR. Rayleigh and related resolution criteria implicitly assume zero noise, and therefore can be viewed as a noise-free limit arising from a more general principle linking resolution with noise.  

Neifeld writes that a ``meaningful definition of resolution should ... depend on the signal-to-noise ratio (SNR)'' \cite{13}. Examples include statistical model-based resolution \cite{33} which invokes the Cram\'{e}r-Rao lower bound \cite{15,16}, the estimation of ghost-imaging resolution via the auto-covariance of a noisy signal \cite{29,30} or via Fourier ring correlation \cite{27}, the use of non-classical states of light such as squeezed light to alter the noise properties of the quantized electromagnetic field so as to improve resolution in certain contexts \cite{14,18}, approaches based on decision theory \cite{12}, information theory \cite{34}, catastrophe theory \cite{35} etc.  These exemplify the link between resolution and noise, and clarify the fact that resolution is intimately related to any {\em a priori} knowledge one may choose to bring---or be able to bring---to a given imaging scenario \cite{4}. For further formulations of resolution, several surveys are recommended \cite{12,17,36}.   

\subsection{Intrinsic imaging quality} 

We now turn to a recent example in which the authors of the present review have been involved.  In intensity-linear imaging systems a trade-off between SR and SNR always exists, and the two cannot be improved simultaneously beyond a certain well-defined limit. This limit depends only on the number of imaging quanta used for image formation \cite{37,38,39,40}. As a result of this principle, a {\em fundamental} characteristic is represented not by SR or SNR alone, but by their dimensionless combination known as ``intrinsic imaging quality''.   

Intrinsic imaging quality $Q_S$ is defined as the following ratio of SNR to SR, normalized by incident fluence:   
\begin{equation}
    Q_S^2=\frac{\textrm{SNR}^2}{(\textrm{SR}_{\textrm{PSF}})^dF_{\textrm{in}}}.
\end{equation}
Here, $d$ is the dimensionality of the image ($d=2$ for conventional planar images) and $F_{\textrm{in}}$ is the imaging particle fluence expressed as the number of registered particles per unit area (or, more generally, the $d$-dimensional volume) of the image \cite{37}.   

We now motivate this definition for $Q_S$, which in $d=2$ dimensions is the SNR per unit SR per unit square-root fluence, as an intrinsic measure of image quality. Suppose $N$ point-like classical imaging quanta, shown as black dots in Fig.~4, fall randomly over a 2D detector with area $A$ (shown in red). Suppose we are free to pixelate this area $A$ with $M$ pixels, each of  equal area, with $M$ being any positive integer.  These pixels, indicated by green squares, are used to bin the point-like classical imaging quanta (e.g.~``photons'', in the loose sense of the term) falling upon $A$. The number of quanta, binned in each pixel, constitute the ``signal'' in each pixel.  

\begin{figure}[h!]
\centering\includegraphics[width=8cm]{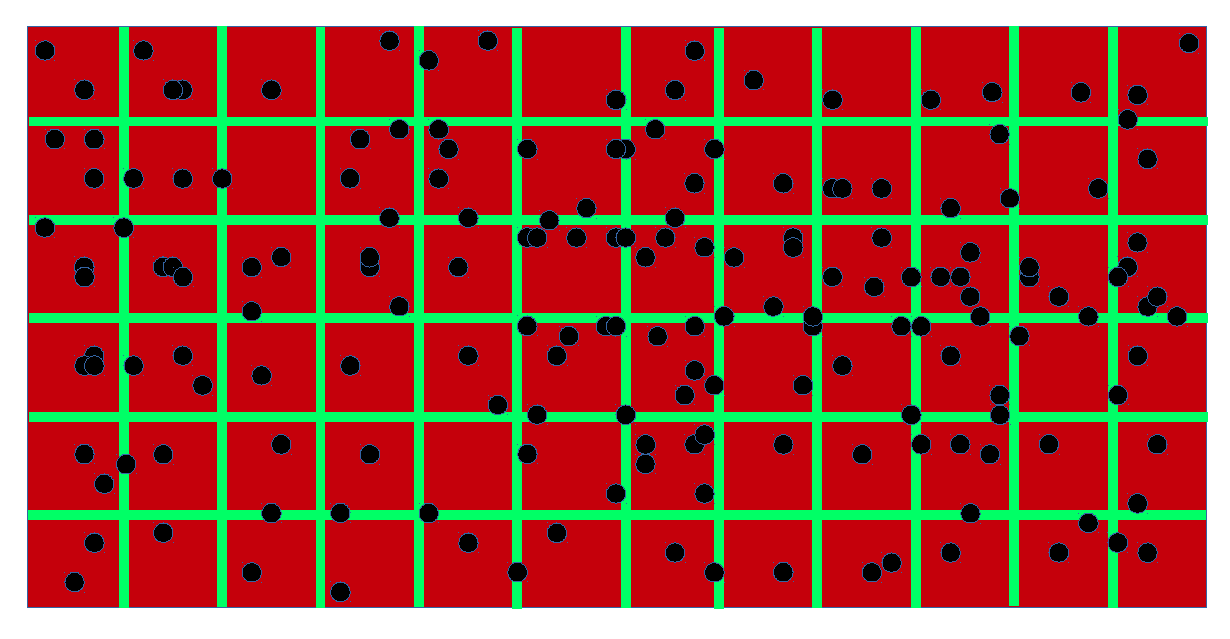}
\caption{AN INTRINSIC MEASURE OF IMAGE QUALITY.  A given number $N$ of imaging quanta is registered as points, shown in black, over an area $A$ shown in red.  This area is divided into $M$ squares, each of which have equal area.  The smaller the size of the green squares, the finer the resolution but the higher the noise of the signal detected within each square.  This is the essence of the noise--resolution tradeoff. }
\end{figure}

(i) Incident fluence $F_{\textrm{in}}$ is the  number of imaging quanta illuminating the detector per unit area, so that $F_{\textrm{in}} = N /A$.  (ii) The spatial resolution $\textrm{SR}_{\textrm{PSF}}$ follows from the fact that $A = M (\textrm{SR}_{\textrm{PSF}})^2$, so that $\textrm{SR}_{\textrm{PSF}} = \sqrt{A/M}$. (iii) The maximum classical SNR per pixel, $\textrm{SNR}_{\textrm{max}}$, is given by Poisson statistics as the square root of the number of quanta per pixel, so $\textrm{SNR} \le \textrm{SNR}_{\textrm{max}} = \sqrt{N/M}$.  This leads us to define the intrinsic quality $Q_S$ as the ratio of the actual SNR to its maximum value, hence $Q_S = \textrm{SNR}/\textrm{SNR}_{\textrm{max}} \le 1$, so that $\textrm{SNR} = Q_S \sqrt{N/M}$. 

If both $M$ and $N$ are eliminated from the estimates for $F_{\textrm{in}}$, $\textrm{SR}_{\textrm{PSF}}$ and $\textrm{SNR}$ given in (i, ii, iii) above, one obtains the $d=2$ version of the definition for $Q_S$ in Eq.~(5).  $Q_S$ is a natural, dimensionless, invariant, intrinsic property of the system which is formed from a combination of the non-system-invariant quantities $\textrm{SNR}$, $\textrm{SR}_{\textrm{PSF}}$, and $F_{\textrm{in}}$.

We have seen, using a heuristic argument centered around Fig.~4, that the fact that a real system cannot improve SNR implies $Q_S \le 1$.  A rigorous argument gives:  
\begin{equation}
    Q_S \le C_d,
\end{equation}
where $C_d$ is a dimensionless constant on the order of unity, which depends only on the dimensionality $d$ of the image \cite{38}. This shows that SNR and SR cannot be improved simultaneously beyond a certain well-defined limit, unless the illumination fluence is increased or {\em a priori} information is used (note that an increase in SR is synonymous with worsened spatial resolution).  On the other hand, it has been also shown that a trade-off between SNR and SR is always possible e.g.~by means of simple linear filtering of the image, which typically increases SNR, but spoils SR, or deconvolution, which does the opposite, i.e.~improves the SR, but lowers the SNR. In view of this trade-off we again see that it is meaningless to consider the SR and SNR in isolation, particularly when it is possible to process the collected image digitally and change either of the two characteristics at will. It is still meaningful to talk about $Q_S$, as this dimensionless parameter is invariant at least with respect to arbitrary linear transformations of the image.  

The tradeoff, due to the inequality $Q_S \le C_d$, is fundamentally different to that implied by the position--momentum Heisenberg uncertainty principle. The difference between Heisenberg and noise--resolution uncertainty principles can be explained as follows. The position--momentum form of the Heisenberg uncertainty principle places a lower limit on the product of the position uncertainty and the associated momentum uncertainty \cite{7}.  In optical terms, this corresponds to a lower limit on the product of (i) spatial resolution due to a single pixel and (ii) the width of the corresponding far-field diffraction pattern that would result if only that particular pixel (or resolution element) were to be illuminated.  This differs from the noise--resolution uncertainty principle $Q_S =\textrm{SNR}/[\textrm{SR}_{\textrm{PSF}}\sqrt{F_{\textrm{in}}}] \le C_2$ (two-dimensional case), which places a lower limit on the spatial resolution $\textrm{SR}_{\textrm{PSF}}$, which is related to the number of imaging quanta that impinge upon the said pixel, and is independent of the width of the corresponding diffraction pattern that would result if only that particular pixel were to have been illuminated.  This is indicative of the more general statement that the noise--resolution and Heisenberg uncertainty principles are independent principles \cite{40}.   

\subsection{Spatial resolution as correlation length}

The fact that $Q_S$ is invariant with respect to arbitrary linear transformations of the image means that SR is proportional to SNR. This arises from the fact that, when a sufficiently narrow linear filter is applied to an image, the PSF is convolved with that filter and the noise autocorrelation function is convolved with the same filter twice \cite{41}, while the signal (being equal to the mean value of intensity within a flat area of the image) essentially does not change \cite{39}. As a consequence, it is also possible to define the SR as a value proportional to SNR: 
\begin{equation}
    \textrm{SR}_{\textrm{correlation}}=(\textrm{SNR}/F_{\textrm{in}})^{2/d},
\end{equation}
where we have implicitly set $Q_S = 1$. It can be shown that such a definition of the SR, which effectively makes it equal to the correlation length of noise in the image (assuming that in the ``ideal'' image the photon noise is uncorrelated, which is often a good approximation), produces quantitative values which are close to the standard deviation of the imaging system's PSF, i.e.~close to $\textrm{SR}_{\textrm{PSF}}$ defined above, when the SR is dominated by the detector resolution. The advantage of this last definition of the SR is in the fact that it is intrinsically linked to SNR and automatically ensures that the trade-off between the SR and SNR is accounted for. It also provides a convenient method for measuring SR in images without the need to use any test resolution patterns or other specialized techniques, such as e.g.~``knife edge'' imaging \cite{42}. Indeed, it is possible to measure $\textrm{SR}_{\textrm{correlation}}$ of the imaging system by simply collecting a flat (featureless) image and calculating its noise power spectrum \cite{43}. This technique is often used for evaluating the performance of pixelated imaging detectors (e.g.~CCDs), where it is further refined via use of the modulation transfer function (MTF) and detector quantum efficiency (DQE). The MTF of an imaging system is the modulus of the Fourier transform of the PSF, while DQE can be viewed as a generalization of $Q_S$. More precisely, DQE can usually be identified with the ratio of the output and input SNR of the imaging system (i.e.~a detector) at a certain spatial Fourier frequency \cite{44}. In the same sense, $Q_S$ is equal to the ratio of these SNRs at the zero frequency, i.e.~on average \cite{39}. Note that $\textrm{SR}_{\textrm{correlation}}$ can naturally include effects of both the first (optical) and the second (digital) stages of a computational imaging system. If, for example, the second step includes some filtering of the raw collected image, it will likely improve SNR, while increasing (worsening) SR in direct agreement with the definition of $\textrm{SR}_{\textrm{correlation}}$. 

However, the disadvantage of the definition of $\textrm{SR}_{\textrm{correlation}}$ lies in the fact that it relies on Poisson statistics of the detected image and also typically cannot account for effects of source size and other pre-detection factors on the SR. Indeed, the size of an incoherent source usually does not affect the correlation length of photon shot noise \cite{41}. Instead, the source size is related to the correlation length of the so-called ``self noise'' of the radiation incident on the detector \cite{41}, but the variance of the latter noise is scaled in proportion to the ratio of the coherence time to the exposure time, which is typically tiny (smaller than $10^{-10}$) in most imaging situations. Therefore, the self-noise does not appreciably contribute to the image noise and the image noise is dominated by the photon shot (detection) noise, which is sensitive to the detector PSF, but is insensitive to the source size.  

\section{Computational-imaging definition of spatial resolution}

To overcome shortcomings of the Rayleigh and related definitions of SR and bring them into the ACI age, apply a methodology somewhat analogous to one used in quantum theory \cite{7}. Consider any image acquisition as a superposition of elementary measurements corresponding to projections of the observable quantity (image intensity) onto a certain state vector. To appreciate how this relates to the case of conventional image detection with a pixelated detector, associate the measured intensity value in each pixel of the detector with the scalar product (i.e.~the integral of the ordinary product) of the image intensity and the ``state'' function, which is unity inside the pixel and zero elsewhere. A conventional image is equivalent to a set of such ``elementary'' single-pixel measurements. The power of such an approach lies in its ability to easily and uniformly incorporate many types of more complicated image measurements, such as e.g.~those used in computational ghost imaging with illumination masks or structured illumination and a single-pixel (``bucket'') detector \cite{2, 45}. Thus the set of all computational imaging systems is a super-set of the set of all conventional imaging systems. 

As soon as one adopts this more general view on the imaging process, it becomes possible to re-define SR in terms of the dimensionality of the space spanned by all probed measurement vectors \cite{46}. In the case of images collected with a pixelated detector, this dimensionality corresponds to the number $M$ of pixels in which the intensity distribution of the image was measured. More generally and precisely, the current approach suggests taking the $d$-dimensional volume, $V$, of the ``image'' and dividing it by the dimensionality, $M$, of the measurement space, to give a computational-imaging SR: 
\begin{equation}
    \textrm{SR}_{\textrm{CI}}=(V/M)^{1/d}.
\end{equation}
An important requirement implied by the term ``dimensionality'' is that the state vectors, onto which the measurements project the intensity distribution, must be linearly independent. It then turns out that if different measurement vectors are not orthogonal, the data produced by the set of such measurements may not be as ``well determined'' as in the case of the same number of measurements made with respect to orthogonal states. Rather than attempting to incorporate this variable degree of ``determinacy'' into the notion of SR, it is more natural to let it affect the associated SNR. Consider for example a reasonable expectation that multiple measurements with respect to linearly dependent state vectors do not increase the SR, but can increase SNR, provided the mean signal grows faster than its variance (i.e., the noise) in the corresponding averaging process. This is true in the case of Poisson statistics and measurements performed with a pixelated detector, where the squared-SNR equals the mean total number of photons registered in each pixel, $N/M$. Now take into account the fact that the incident photon fluence is equal to the number of registered photons divided by the volume of the image, i.e.~$F_{\textrm{in}}=N/V$. Using this definition, we obtain  
\begin{equation}
    \tilde{Q}_S^2=\frac{\textrm{SNR}^2}{(\textrm{SR}_{\textrm{CI}})^d F_{\textrm{in}}}=\frac{(N/M)}{(V/M)(N/V)}=1.
\end{equation}
The modified ``imaging quality'', $\tilde{Q}_S$, is defined here with respect to the new definition of the spatial resolution, $\textrm{SR}_{\textrm{CI}}$. The invariance of this $\tilde{Q}_S$ occurs in the case of Poisson statistics, when the spatial resolution is precisely proportional to the SNR. For other types of photon statistic, e.g.~Gaussian or sub-Poissonian, $\tilde{Q}_S$ may either decrease or increase with the increase of the number of registered photons. The latter could be the result of  increased incident fluence, increased exposure time or increased number of repeated measurements. As suggested above, SNR may also be lower than $N/M$, if the $M$ measurements are performed with respect to linearly independent, but not orthogonal state vectors \cite{46}. For example, it is possible to show \cite{46} that in the case of repeated measurements with sub-pixel shifts of the detector (or the object) between the measurements, as sometimes practiced in super-resolution microscopy, the squared-SNR behaves like $N/M^2$, and so the imaging quality progressively deteriorates with an increased number of measurements: $\tilde{Q}_S=1/M$. In such cases, similar results indicate that the imaging particles are not used as efficiently as in the case of orthogonal measurements, so that the same radiation dose delivered to the sample produces less information about the sample. Alternatively, using orthogonal measurement vectors and entangled photons for illumination may lead to sub-Poisson statistics in the image and values of $\tilde{Q}_S > 1$.  

\section{Conclusions}

Historical notions of SR and SNR in optics and imaging warrant revision in the computational imaging era. The multi-stage hybrid process involving analog and digital processing of collected image data {\em en route} to final images is typical for most modern imaging and optical images. Old definitions of ``raw'' spatial resolution are increasingly incompatible with this process of image formation and need to be modernized accordingly. We suggested several possible working definitions of SR and presented theoretical and practical arguments explaining reasons for their introduction and advantages associated with their use. We emphasized the intrinsic duality between SR and SNR and the need to consider the two  together to obtain meaningful and workable definitions consistent with image processing operations utilized in computational imaging. Finally, we discussed how the generic ideas of quantum metrology could be utilized to formulate the most general, but still practically useful, definitions of SR, SNR and imaging quality. We gave examples drawn from super-resolution optical reconstruction microscopy, holography, tomography and ghost imaging. 

\bigskip

\noindent {\em Note: A preliminary version of this article was presented at the SPIE Photonics West 2019 conference in San Francisco, U.S.A., held on February 2-5, 2019.  This was published as T. E. Gureyev, D. M. Paganin, A. Kozlov, and H. M. Quiney ``Spatial resolution and signal-to-noise ratio in x-ray imaging'', Proc. SPIE {\bf 10887}, Quantitative Phase Imaging V, 108870J (4 March 2019).} 


\section*{Acknowledgments}

The authors acknowledge useful discussions with Christian Dwyer and Harry Quiney.


\begin{thebibliography}{0}%
\makeatletter
\providecommand \@ifxundefined [1]{%
 \@ifx{#1\undefined}
}%
\providecommand \@ifnum [1]{%
 \ifnum #1\expandafter \@firstoftwo
 \else \expandafter \@secondoftwo
 \fi
}%
\providecommand \@ifx [1]{%
 \ifx #1\expandafter \@firstoftwo
 \else \expandafter \@secondoftwo
 \fi
}%
\providecommand \natexlab [1]{#1}%
\providecommand \enquote  [1]{``#1''}%
\providecommand \bibnamefont  [1]{#1}%
\providecommand \bibfnamefont [1]{#1}%
\providecommand \citenamefont [1]{#1}%
\providecommand \href@noop [0]{\@secondoftwo}%
\providecommand \href [0]{\begingroup \@sanitize@url \@href}%
\providecommand \@href[1]{\@@startlink{#1}\@@href}%
\providecommand \@@href[1]{\endgroup#1\@@endlink}%
\providecommand \@sanitize@url [0]{\catcode `\\12\catcode `\$12\catcode
  `\&12\catcode `\#12\catcode `\^12\catcode `\_12\catcode `\%12\relax}%
\providecommand \@@startlink[1]{}%
\providecommand \@@endlink[0]{}%
\providecommand \url  [0]{\begingroup\@sanitize@url \@url }%
\providecommand \@url [1]{\endgroup\@href {#1}{\urlprefix }}%
\providecommand \urlprefix  [0]{URL }%
\providecommand \Eprint [0]{\href }%
\providecommand \doibase [0]{https://doi.org/}%
\providecommand \selectlanguage [0]{\@gobble}%
\providecommand \bibinfo  [0]{\@secondoftwo}%
\providecommand \bibfield  [0]{\@secondoftwo}%
\providecommand \translation [1]{[#1]}%
\providecommand \BibitemOpen [0]{}%
\providecommand \bibitemStop [0]{}%
\providecommand \bibitemNoStop [0]{.\EOS\space}%
\providecommand \EOS [0]{\spacefactor3000\relax}%
\providecommand \BibitemShut  [1]{\csname bibitem#1\endcsname}%
\let\auto@bib@innerbib\@empty
\end{thebibliography}%


\begin{thebibliography}{1}
\newcommand{\enquote}[1]{``#1''}

\bibitem{1}
D.~Gabor,
\enquote{A new microscopic principle,}
{ {Nature}} \textbf{161}, 777--778 (1948).

\bibitem{2}
J.~N.~Mait, G.~W.~Euliss and R.~A.~Athale,
\enquote{Computational Imaging,}
{ {Adv. Opt. Phot.}} \textbf{10}, 409--483 (2018).

\bibitem{3}
Lord~Rayleigh,
\enquote{Investigations in optics, with special reference to the spectroscope,}
{ {Phil. Mag.}} \textbf{8}, 261--274 (1879).

\bibitem{4}
G.~Toraldo~di~Francia,
\enquote{Resolving power and information,}
{ {J. Opt. Soc. Am.}} \textbf{45}, 497--501 (1955).

\bibitem{5}
A.~Papoulis,
\enquote{Systems and Transforms with Applications in Optics}
(Robert E. Krieger Publishing Company, 1981).

\bibitem{6}
M.~Born and E.~Wolf,
\enquote{Principles of Optics, 7th ed.}
(Cambridge University Press, 1999).

\bibitem{8}
V.~Ronchi,
\enquote{Resolving power of calculated and detected images,}
{ {J. Opt. Soc. Am.}} \textbf{51}, 458--460 (1961).

\bibitem{9}
N.~J.~Bershad,
\enquote{Resolution, optical-channel capacity and information theory,}
{ {J. Opt. Soc. Am.}} \textbf{59}, 157--163 (1969).

\bibitem{10a}
D.~L.~Fried,
\enquote{Resolution, signal-to-noise ratio, and measurement precision,}
{ {J. Opt. Soc. Am.}} \textbf{69}, 399--406 (1979).

\bibitem{10b}
D.~L.~Fried,
\enquote{Resolution, signal-to-noise ratio, and measurement precision: addendum}
{ {J. Opt. Soc. Am.}} \textbf{70}, 748--749 (1980).

\bibitem{11}
I.~J.~Cox and C.~J.~R.~Sheppard,
\enquote{Information capacity and resolution in an optical system,}
{ {J. Opt. Soc. Am. A}} \textbf{3}, 1152--1158 (1986).

\bibitem{12}
A.~J.~den~Dekker and A.~van~den~Bos,
\enquote{Resolution: a survey,}
{ {J. Opt. Soc. Am. A}} \textbf{14}, 547--557 (1997).

\bibitem{13}
M.~A.~Neifeld,
\enquote{Information, resolution, and space--bandwidth product,}
{ {Opt. Lett.}} \textbf{23}, 1477--1479 (1998).

\bibitem{14}
M.~I.~Kolobov and C.~Fabre,
\enquote{Quantum limits on optical resolution,}
{ {Phys. Rev. Lett.}} \textbf{85}, 3789--3792 (2000).

\bibitem{15}
S.~Ram, E.~S.~Ward, and R.~J.~Ober,
\enquote{Beyond Rayleigh's criterion: A resolution measure with application to single-molecule microscopy,}
{ {Proc. Natl Acad. Sci. U. S. A.}} \textbf{103}, 4457--4462 (2006).

\bibitem{16}
A.~Amar and A.~J.~Weiss,
\enquote{Fundamental limitations on the resolution of deterministic signals,}
{ {IEEE Trans. Sign. Proc.}} \textbf{56}, 5309--5318 (2008).

\bibitem{17}
M. Sato,
\enquote{Resolution,}
in \enquote{Handbook of Charged Particle Optics, 2nd ed.}  J. Orloff, ed., Chap. 20, pp.~391-435 (CRC Press, 2008).

\bibitem{18}
W.~Larson and B.~E.~A.~Saleh,
\enquote{Resurgence of Rayleigh's curse in the presence of partial coherence,}
{ {Optica}} \textbf{5}, 1382--1389 (2018).

\bibitem{7}
A.~Messiah,
\enquote{Quantum Mechanics, Vol.~1}
(North-Holland, 1961).

\bibitem{19}
L.~Yaroslavsky and M.~Eden,
\enquote{Fundamentals of Digital Optics: Digital Signal Processing in Optics and Holography}
(Birkhauser, 1996). 

\bibitem{20}
J.~Miao, P.~Charalambous, J.~Kirz, and D.~Sayre,
\enquote{Extending the methodology of X-ray crystallography to allow imaging of micrometre-sized non-crystalline specimens,}
{ {Nature}} \textbf{400}, 342--344 (1999).

\bibitem{21}
F.~Natterer,
\enquote{The Mathematics of Computerised Tomography}
(John Wiley \& Sons and Teubner, 1986). 

\bibitem{22}
B.~J.~Hoenders,
\enquote{Review of a bewildering classical-quantum phenomenon: Ghost imaging,}
{ {Adv. Imag. Elect. Physics}} \textbf{208}, 1--41 (2018).

\bibitem{23}
D.~Paganin, T.~E.~Gureyev, S.~C.~Mayo, A.~W.~Stevenson, Ya.~I.~Nesterets, and S.~W.~Wilkins,
\enquote{X-ray omni microscopy,}
{ {J. Microsc.}} \textbf{214}, 315--327 (2004).

\bibitem{24}
M.~Rust, M.~Bates, and X.~Zhuang,
\enquote{Sub-diffraction-limit imaging by stochastic optical reconstruction microscopy (STORM),}
{ {Nat. Meth.}} \textbf{3}, 793--796 (2006).

\bibitem{25}
H.~Deschout, F.~Cella~Zanacchi, M.~Mlodzianoski, A.~Diaspro, J.~Bewersdorf, S.~T.~Hess, and K.~Braeckmans,
\enquote{Precisely and accurately localizing single emitters in fluorescence microscopy,}
{ {Nat. Meth.}} \textbf{11}, 253--266 (2014).

\bibitem{26}
J.~Vangindertael, R.~Camacho, W.~Sempels, H.~Mizuno, P.~Dedecker, and K.~P.~F.~Janssen,
\enquote{An introduction to optical super-resolution microscopy for the adventurous biologist,}
{ {Methods Appl. Fluoresc.}} \textbf{6}, 022003 (2018).

\bibitem{27}
A.~M.~Kingston, D.~Pelliccia, A.~Rack, M.~P.~Olbinado, Y.~Cheng, G.~R.~Myers, and D.~M.~Paganin,
\enquote{Ghost tomography,}
{ {Optica}} \textbf{5}, 1516--1520 (2018).

\bibitem{28}
A.~M.~Kingston, G.~R.~Myers, D.~Pelliccia, I.~D.~Svalbe, and D.~M.~Paganin,
\enquote{X-ray ghost-tomography: Artefacts, dose distribution and mask considerations,}
{ {IEEE Trans. Comput. Imaging}} \textbf{5}, 136--149 (2019).

\bibitem{29}
F.~Ferri, D.~Magatti, L.~A.~Lugiato, and A.~Gatti,
\enquote{Differential ghost imaging,}
{ {Phys. Rev. Lett.}} \textbf{104}, 253603 (2010).

\bibitem{30}
D.~Pelliccia, M.~P.~Olbinado,  A.~Rack,  A.~M.~Kingston,  G.~R.~Myers, and  D.~M.~Paganin,
\enquote{Towards a practical implementation of x-ray ghost imaging with synchrotron light,}
{ {IUCrJ}} \textbf{5}, 428--438 (2018).

\bibitem{31}
J.~Nolte,
\enquote{The Human Brain: An Introduction to its Functional Anatomy, 6th ed.}
(Mosby Elsevier, 2009). 

\bibitem{32}
C.~C.~Aggarwal,
\enquote{Neural Networks and Deep Learning}
(Springer International, 2018). 

\bibitem{33}
S.~Van~Aert, A.~J.~den~Dekker, D.~Van~Dyck, and A.~Van~den~Bos,
\enquote{The Notion of Resolution,}
in \enquote{Science of Microscopy,}  P.~W.~Hawkes and J.~C.~H.~Spence eds., pp. 1228--1265 (Springer, 2007).

\bibitem{34}
L.~Motka, B.~Stoklasa, M.~D'Angelo, P.~Facchi, A.~Garuccio, Z.~Hradil, S.~Pascazio, F.~V.~Pepe, Y.~S.~Teo, J.~Rehacek, and L.~L.~Sanchez-Soto,
\enquote{Optical resolution from Fisher information,}
{ {Eur. Phys. J. Plus}} \textbf{131}, 130 (2016).

\bibitem{35}
A.~van~den~Bos,
\enquote{Optical resolution: an analysis based on catastrophe theory,}
{ {J. Opt. Soc. Am. A}} \textbf{4}, 1402--1406 (1987).

\bibitem{36}
A.~van~den~Bos and A.~J.~den~Dekker,
\enquote{Resolution reconsidered - Conventional approaches and an alternative,}
{ {Adv. Imaging Elect. Phys.}} \textbf{117}, 241--360 (2001).

\bibitem{37}
T.~E.~Gureyev, Ya.~I.~Nesterets, F.~de~Hoog, G.~Schmalz, S.~C.~Mayo, S.~Mohammadi, and G.~Tromba,
\enquote{Duality between noise and spatial resolution in linear systems,}
{ {Opt. Express}} \textbf{22}, 9087--9094 (2014).

\bibitem{38}
F.~de~Hoog, G.~Schmalz, and T.~E.~Gureyev,
\enquote{An uncertainty inequality,}
{ {Appl. Math. Lett.}} \textbf{38}, 84--86 (2014).

\bibitem{39}
T.~E.~Gureyev, Ya.~I.~Nesterets, and F.~de~Hoog,
\enquote{Spatial resolution, signal-to-noise and information capacity of linear imaging systems,}
{ {Opt. Express}} \textbf{24}, 17168--17182 (2016).

\bibitem{40}
T.~E.~Gureyev, F.~de~Hoog, Ya.~I.~Nesterets, and D.~M.~Paganin,
\enquote{On the noise-resolution duality, Heisenberg uncertainty and Shannon's information,}
{ {ANZIAM J.}} \textbf{56}, C1--C5 (2015).

\bibitem{41}
J.~W.~Goodman,
\enquote{Statistical Optics}
(John Wiley \& Sons, 2000). 

\bibitem{42}
J.~M.~Heck, D.~T.~Attwood, W.~Meyer-Ilse, and E.~H.~Anderson,
\enquote{Resolution determination in X-ray microscopy: an analysis of the effects of partial coherence and illumination spectrum,}
{ {J. X-Ray Sci. Tech.}} \textbf{8}, 94--104 (1998).

\bibitem{43}
P.~Baran, S.~Pacile, Ya.~I.~Nesterets, S.~C.~Mayo, C.~Dullin, D.~Dreossi, F.~Arfelli, D.~Thompson, D.~Lockie, M.~McCormack, S.~T.~Taba, F.~Brun, M.~Pinamonti, C.~Nickson, C.~Hall, M.~Dimmock, F.~Zanconati, M.~Cholewa, H.~Quiney, P.~C.~Brennan, G.~Tromba, and T.~E.~Gureyev, \enquote{Optimization of propagation-based x-ray phase-contrast tomography for breast cancer imaging,}
{ {Phys. Med. Biol.}} \textbf{62}, 2315--2332 (2017).

\bibitem{44}
I.~A.~Cunningham and R.~Shaw,
\enquote{Signal-to-noise optimization of medical imaging systems,}
{ {J. Opt. Soc. Am. A}} \textbf{16}, 621--632 (1999).

\bibitem{45}
P.-A.~Moreau, E.~Toninelli, T.~Gregory, and M.~J.~Padgett,
\enquote{Ghost imaging using optical correlations,}
{ {Laser Photonics Rev.}} \textbf{12}, 1700143 (2017).

\bibitem{46}
T.~E.~Gureyev, D.~M.~Paganin, A.~Kozlov, Ya.~I.~Nesterets, and H.~M.~Quiney,
\enquote{Complementary aspects of spatial resolution and signal-to-noise ratio in computational imaging,}
{ {Phys. Rev. A}} \textbf{97}, 053819 (2018).






 \end{thebibliography}
\end{document}